\documentclass[american,english,journal=jctce,manuscript=article,layout=twocolumn]{achemso}
    \usepackage{refstyle}
    \usepackage[T1]{fontenc}
    \usepackage{palatino}
    \usepackage{multirow}
    \usepackage{graphicx}

    \usepackage{adjustbox} %
    \usepackage{xcolor} %
    \usepackage{amsmath} %
    \usepackage{amssymb} %
    \usepackage{amsbsy} %
    \usepackage{bm} %
    \usepackage{fixmath}
    \usepackage{textcomp} %
    \usepackage{amsfonts,mathrsfs}

    \makeatletter
    \title{Predicting the kinetics of RNA oligonucleotides using Markov state models}

    \author{Giovanni Pinamonti}
    \affiliation{Scuola Internazionale Superiore di Studi Avanzati, 
International School for Advanced Studies,
265, Via Bonomea I-34136 Trieste, Italy}
    \altaffiliation{Joint first authorship}

    \author{Jianbo Zhao}
    \affiliation{Department of Chemistry, University of Rochester, Rochester, New York 14627, United States}
    \altaffiliation{Joint first authorship}

    \author{David E. Condon}
    \affiliation{Department of Chemistry, University of Rochester, Rochester, New York 14627, United States}

    \author{Fabian Paul}
    \affiliation{Department for Mathematics and Computer Science, Freie Universit{\"a}t, Arnimallee 6, Berlin 14195, Germany}

    \author{Frank No\'e}
    \affiliation{Department for Mathematics and Computer Science, Freie Universit{\"a}t, Arnimallee 6, Berlin 14195, Germany}

    \author{Douglas H. Turner}
    \affiliation{Department of Chemistry, University of Rochester, Rochester, New York 14627, United States}

    \author{Giovanni Bussi}
    \affiliation{Scuola Internazionale Superiore di Studi Avanzati, 
International School for Advanced Studies,
265, Via Bonomea I-34136 Trieste, Italy}
    \email{bussi@sissa.it}               %

    \newcommand{\vect}[1]{\boldsymbol{#1}}

\newcommand*{\sometext}{

\textbf{
Nowadays different experimental techniques, such as single molecule or 
relaxation experiments, can provide dynamic properties of biomolecular systems, 
but the amount of detail obtainable with these methods is often limited 
in terms of time or spatial resolution.
Here we use state-of-the-art computational techniques, namely atomistic
molecular dynamics and Markov state models, to provide insight into the rapid
dynamics of short RNA oligonucleotides, 
in order to elucidate the kinetics of stacking interactions.
Analysis of multiple microsecond-long simulations indicates that the main 
relaxation modes of such molecules can consist of transitions 
between alternative folded states,
rather than between random coils and native structures.
After properly removing structures that are artificially stabilized
by known inaccuracies of the current RNA AMBER force field,
the kinetic properties 
predicted
are consistent with the timescales
of previously reported relaxation experiments.
}
 }
\let\oldmaketitle\maketitle
\let\maketitle\relax

\begin{document}

    \noindent \twocolumn[
      \begin{@twocolumnfalse}
Reprinted with permission from:\\
``Predicting the kinetics of RNA oligonucleotides using Markov state models''\\
Journal of Chemical Theory and Computation.\\
doi: 10.1021/acs.jctc.6b00982. Copyright 2016 American Chemical Society.\\
      \end{@twocolumnfalse}
    ]
    \maketitle
    \makeatother

    \noindent \twocolumn[
      \begin{@twocolumnfalse}
        \oldmaketitle
        \begin{abstract}
          \sometext
          \leavevmode \\
        \end{abstract}
      \end{@twocolumnfalse}
    ]

    \section{Introduction}

The importance of ribonucleic acid (RNA) in molecular biology is constantly growing, as researchers discover new roles played by non-coding RNAs
in the cell.\cite{morris2014riseregRNA}
In many cases RNA function relies on complex multi-step conformational transitions that occur in response to cellular signals,\cite{al2008rna}
calling for an effort in determining not only
RNA structure but also RNA dynamics.
RNA stability depends on a large variety of interactions, 
including stacking, hydrogen bonding, 
and interactions with water and ions~\cite{bloomfield2000nucleic}.
In vacuum, stacking interactions arise from %
complex interactions between 
aromatic rings~\cite{hobza1999structure}.
However, in biological environment these interactions are heavily mediated by water.
Dinucleotides and short oligonucleotides are perfect models to study stacking in RNA.
While the equilibrium properties have been extensively characterized by NMR measurements~\cite{lee1976conformational,ezra1977conformational,lee1980conformation,olsthoorn1982influence,lee1983conformational,vokacova2009structure,yildirim2011benchmarking,tubbs2013nuclear,condon2015stacking}
their kinetics have been only studied in a limited number of temperature-jump (T-jump) experiments~\cite{porschke1976cable,porschke1978molecular,dewey1979laser}. 
Molecular dynamics (MD) provides a tool that can be used to characterize in detail the time evolution of these systems
at picosecond and Angstrom resolution,
and can supply insightful fine-detailed information that can complement experimental measurements~\cite{sponer2014molecular}.
Several MD studies on RNA oligonucleotides have been published to date (see, \emph{e.g.},
refs.~\cite{vokacova2009structure,yildirim2011benchmarking,bergonzo2013multidimensional,condon2015stacking,bergonzo2015highly,Gil-LeyBussi2015,brown2015stacking,gil2016empirical,cesari2016combining}).
MD was also used to characterize the stacking free energy
associated with base fraying\cite{colizzi2012rna}
and the stacking thermodynamics in DNA.\cite{hase2016free}
However, all these works were focused on the equilibrium properties rather than on the relaxation
times.
Kinetics of RNA tetraloops have been investigated\cite{bowman2008structural}, but the complexity
of the system led to unconverged results.

In this work %
we present a systematic analysis 
of the kinetic processes for RNA oligonucleotides, 
as predicted by MD simulations.
We used Markov state models (MSMs) and hidden Markov models (HMMs), 
to provide a complete description of the
transitions characterized by the slowest relaxation times. 
We studied a number of dinucleoside monophosphates,
a trinucleotide (AAA), and a tetranucleotide (AAAA) so as to characterize the dependence of kinetics
on length and sequence. 
Results are compared with available experiments. Whereas 
some of the reported transitions correspond to known artifacts of the current force field,
our results can explain the overall trends. Importantly, we suggest that
measured autocorrelation times may not be directly 
associated to transitions between helix and coil structures
but to transitions between kinetic traps characterized by different stacking patterns.

    \section{Methods}

    \subsection{Molecular dynamics simulations}

MD simulations were run with different salt concentrations, ionic strength, 
sequence, and oligonucleotide length. 
The dinucleotides and trinucleotide simulations were performed using
GROMACS 4.6.7~\cite{pronk2013gromacs}.
The tetranucleotide simulation was run using 
AMBER 11~\cite{case2011amber}. 
We used AMBER force-field parameters~\cite{hornak2006comparison}
with parmbsc0~\cite{perez2007refinement}
and $\chi$OL3~\cite{zgarbova2011refinement}
corrections. %
Trajectories of di- and tri-nucleotides were generated in the isothermal-isobaric
ensemble using stochastic velocity rescaling~\cite{bussi2007canonical} 
and the Parrinello-Rahman barostat~\cite{parrinello1981polymorphic}.
Thermostat and barostat used in AAAA simulations are described 
in~\cite{condon2015stacking}.
RNA molecules were solvated in 
 explicit water
(TIP3P parameters~\cite{jorgensen1983comparison}), adding Na$^+$ counterions to neutralize the RNA
charge, plus additional NaCl to reach the nominal concentration.
Details of all simulations are reported in Tab.~\ref{tab:MD}.
For each of the studied systems we ran several MD simulations, starting from different initial configurations,
and stored on the disk frames with the time stride indicated in Tab~\ref{tab:MD}.

\begin{table*}
\centering
    \begin{tabular}{lcccccccccc}
\hline
\rotatebox{90}{\small  Sequence} &
\rotatebox{90}{\small $T$ (K)} &
\rotatebox{90}{\small Na$^+$ (M)} & 
\rotatebox{90}{\small N. traj.} & 
\rotatebox{90}{\small Total length (\textmu s)} &
\rotatebox{90}{\small Stride (ps)} &
\rotatebox{90}{\small TICA lagtime (ns)} &
\rotatebox{90}{\small TICA dimensions} &
\rotatebox{90}{\small Total N. of microstates} &
\rotatebox{90}{\small MSM lagtime (ns)} &
\rotatebox{90}{\small N. of active microstates}\\
\hline
CC   & $277$ & $1.0 $& $4 $& $9.6 $ & $10 $   & $1.0$ & $10$ & $400$ & $0.5 $ & $367$ \\
AC   & $277$ & $1.0 $& $4 $& $9.7 $ & $10 $   & $1.0$ & $10$ & $400$ & $0.5 $ & $374$ \\
CA   & $277$ & $1.0 $& $4 $& $9.1 $ & $10 $   & $1.0$ & $10$ & $400$ & $0.5 $ & $381$ \\
AA   & $277$ & $1.0 $& $4 $& $8.9 $ & $10 $   & $1.0$ & $10$ & $400$ & $0.5 $ & $397$ \\
CC   & $300$ & $1.0 $& $8 $& $7.0 $ & $10 $   & $1.0$ & $10$ & $400$ & $0.5 $ & $377$ \\
AC   & $300$ & $1.0 $& $8 $& $7.0 $ & $10 $   & $1.0$ & $10$ & $400$ & $0.5 $ & $375$ \\
CA   & $300$ & $1.0 $& $8 $& $6.6 $ & $10 $   & $1.0$ & $10$ & $400$ & $0.5 $ & $383$ \\
AA   & $300$ & $1.0 $& $16$& $7.0 $ & $10 $   & $1.0$ & $10$ & $400$ & $0.5 $ & $398$ \\
AAA  & $300$ & $0.1 $& $17$& $57.0$ & $100$   & $5.0$ & $19$ & $100$ & $5.0 $ & $100$ \\
AAAA & $275$ & $0.13$& $4 $& $35  $ & $100$   & $5.0$ & $44$ & $400$ & $20.0$ & $399$ \\
\hline %
\end{tabular}
\caption{Details of the MD simulations and of the MSM construction.}
\label{tab:MD}
\end{table*}

    \subsection{Markov state model}
MSMs are powerful tools that enable extraction of relevant kinetic information from multiple MD simulations.
See~\cite{noe2008transition,pande2010everything,chodera2014markov} 
for a brief introduction to this topic, 
or~\cite{prinz2011markov,bowman2013introduction}
for a more detailed discussion. Here we summarize the basic concepts which are relevant for the present work.

The idea behind a MSM is to reduce the complexity of an MD simulation 
by dividing the phase space into discrete microstates
(e.g. clustering the frames of the trajectory). 
It is then possible to compute the transition matrix, whose elements,
$T_{ij}$, represent the probability that the system, starting from  
microstate $i$, will transition 
to microstate $j$ after a time, $\tau$.

If these microstates are obtained by a sufficiently fine
discretization of the slow collective coordinates of the
system, powers of this transition matrix can model
the long-time evolution of the dynamics, i.e. the
kinetics and stationary behavior of the system, with
excellent accuracy~\cite{prinz2011markov}.
By performing a spectral analysis of matrix $\vect{T}$ we can
decompose the dynamics of the system into independent processes, each represented by the
$i$th eigenvector of $\vect{T}$~\cite{prinz2011markov}. 
The timescales of such processes can be computed 
from the eigenvalues, $\lambda_i$, of $\vect{T}$ as
\begin{equation}
t_i=-\frac \tau {\ln |\lambda_i|}
\end{equation}

In order to analyze the trajectories produced from the MD we considered the following set of coordinates:
\begin{enumerate}
\item G-vectors (4D vectors connecting the nucleobases ring centers, as described in~\cite{bottaro2014role}) 
\item Backbone dihedrals 
\item Sugar ring torsional angles 
\item Glycosidic torsional angles
\end{enumerate}
The dimensionality of the input data was then reduced using
time-lagged independent components analysis (TICA)~\cite{molgedey1994separation,perez2013identification,schwantes2013improvements}.
Data were projected on the slowest time-lagged independent components (TICs) 
using a kinetic-map projection~\cite{noe2015kinetic}
and then
discretized in microstates using a k-means clustering algorithm~\cite{macqueen1967some}. 
A lag-time $\tau$ was used to construct MSMs that approximate the
dynamics of the discretized systems. Detailed balance was imposed
by constructing reversible MSMs using the procedure described in Ref.~\cite{prinz2011markov}.
Statistical uncertainties were estimated by means of the
Markov chain Monte Carlo sampling of transition
matrices from the posterior distribution, with a reversible prior, as described in~\cite{trendelkamp2015estimation}.
All details and parameters used in the MSMs construction are reported in Tab.~\ref{tab:MD}.
Preliminary tests on selected systems showed that the results of our models do not
vary significantly when changing the values of such parameters (data not shown).
The MSM construction and analysis was performed using the software PyEMMA 2.2~\cite{scherer2015pyemma}.

\subsection{Combined discretization of dinucleotides trajectories}
Since the dinucleotide systems share the same number of residues and the same
backbone, the number of coordinates is the same for all of them. This can be exploited to
perform TICA on a virtual trajectory obtained merging
all the individual trajectories of the dinucleotides.
We discretized the merged trajectories using  k-mean clustering. 
For each dinucleotide system, we then built a separate MSM.
Since not all microstates 
are visited by all the eight systems, each of the resulting MSMs will be 
defined on a subset of the total number of microstates, that we call ``active set''.
From Tab.~\ref{tab:MD} we notice that the fraction of active states is always close to $1$,
indicating that the systems share several common features.

\subsection{Analysis of the kinetics}
From the eigenvectors and the eigenvalues of the transition matrix of an MSM we
can obtain detailed information about the slow processes occurring during the
simulations, as well as a precise estimation of their predicted
timescales.

The eigenvectors of the different dinucleotides' MSMs were then 
compared using an appropriate measure of similarity.
Since the active sets of different MSMs is different we first 
mapped all the eigenvectors, 
$\vect{\psi}$, to a common $400$-dimensional space, defining
\begin{equation}
\tilde{\psi}_i =  
\{ \begin{array}{l}
\sqrt{(p_{eq})_i}\psi_i \quad \text{if} \, i \in A  \\
0 \quad \text{otherwise}  \\
 \end{array} 
\end{equation}
Here, $i$ is the index of the microstate, 
$A$ is the set of active microstates, and $\vect{p}_{eq}$ is the stationary distribution of the MSM considered. 
The eigenvectors are normalized so that $\sum_i \tilde{\psi}_i^2=1$.
We then compute the similarity between two eigenvectors, $\tilde{\vect{\psi}}^\alpha$, $\tilde{\vect{\psi}}^\beta$, from different MSMs  
as the square of their
scalar product, $( \tilde{\vect{\psi}}^{\alpha} \cdot \tilde{\vect{\psi}}^{\beta} )^2$.
We also used kernel principal components analysis~({KPCA})~\cite{scholkopf1997kernel}
to project the first three eigenvectors of the eight 
dinucleotides' MSMs on a 2-D surface, in order to visually group
similar processes from different MSMs. 
As kernel definition we used 
$\Phi(\tilde{\vect{\psi}})=\tilde{\vect{\psi}}\otimes\tilde{\vect{\psi}}$, 
where $\otimes$ denotes the outer product. 
This is invariant for changes in sign of $\vect{\tilde{\psi}}$.
This analysis was possible since the MSMs share a
common set of microstates, given that the clustering was performed on the joint set of MD data of all dinucleotide systems.

As a further analysis of the dinucleotides' slow processes, we computed the
correlations of these eigenvectors with all the dihedral angles of the
dinucleotides. The variables with the highest correlation
coefficient with a given eigenvector
 should be the best suited to describe the correspondent transition
(as explained in~\cite{perez2013identification}).
To avoid ambiguities due to the periodicity
of dihedrals we compute the correlation between eigenvector
$\vect{\psi}$ and torsion $\theta$ as
$\max_\eta [\text{corr}_t(\psi_t,\cos(\theta_t+\eta)) ]$,
where $\psi_t$ is the value of the eigenvector $\vect{\psi}$
on the microstate visited by the system at time $t$, $\theta$ is the value
of the torsion at time $t$, $\text{corr}_t$ indicates the
Pearson product-moment correlation coefficient computed considering all the frames
in the MD trajectories.

The major non-bonded interaction in short oligonucleotides
is the stacking interaction between consecutive nucleobases.
In order to study this we used the stacking definition proposed 
in~\cite{condon2015stacking}, that takes into
account 1) the distance between the centers of mass of the two
nucleobases, 2) the angle defined by the distance vector between the two centers of mass
and the vector normal to the first base plane, 3) the angle between the
two vectors normal to the two bases' planes.
These quantities are combined in a score, $s$, that goes from $-2$ to $+2$. Nucleotides are considered stacked if $s>1$, unstacked otherwise.

The definition used for stacking contains terms 
that affect UV absorption~\cite{cantor1980biophysical}.
For comparing the time dependence of conformational changes
predicted with MD simulations with those measured experimentally,
however, it is not necessary to quantify the change in absorption.
It is only necessary to assume that different states will have
different absorbances.

To further simplify the tri- and tetra-nucleotide models and analyze their features
we used the
kinetic information from the MSMs to lump the microstates into a few
metastable macrostates. This was done using a HMM, as
described in~\cite{noe2013projected}. 
The idea of this method is to model the system
as a Markov chain between a small number of hidden macrostates,
each of which has a different probability distribution
to generate one of the output microstates, which are the ones observed in the simulation.
The parameters that define a HMM are the transition probabilities between hidden states and the probabilities 
of observing each microstate given the current hidden macrostate of the system. 
The optimal values of these parameters can be found via a likelihood-maximization procedure,
that we carried out using the dedicated algorithm included in PyEMMA 2.2~\cite{scherer2015pyemma}.
The resulting metastable macrostates were then analyzed
by looking at the distributions of selected observables (dihedrals,
distances between key atoms, G-vectors~\cite{bottaro2014role}, and stacking score~\cite{condon2015stacking} between bases).

\subsection{Comparison with relaxation experiments}

MSM predictions can be compared with relaxation experiments that probe 
the kinetics of biomolecules. An exhaustive explanation
of the theory behind this comparison is given in Refs.~\cite{noe2011dynamical,buchete2008coarse}. Here we will briefly summarize the key concepts.

Consider a system described by a MSM with $n$ microstates and transition matrix $\vect{T}$.
In a typical relaxation experiment 
a perturbation of the thermodynamic state of the system
(e.g. a change in temperature)
results in the starting distribution, $\vect{p}_0$,
becoming out of equilibrium.
The system then relaxes to its new equilibrium distribution.
The relaxation process is monitored by measuring the evolution of
an observable, $A$, which is a suitable function of the state of
the system.
The time-evolution of $A$ during the relaxation process 
is given by
\begin{equation}
A(t)=A_{eq}+\sum_{i=2}^n \exp{\left(-\frac{t}{t_i}\right)} \gamma_i%
\end{equation}
Where $A_{eq}$ is the value of $A$ at the final equilibrium, and $\gamma_i
$ 
is the amplitude of the $i$th decay process,
which in general depends both on the shape of $\vect{p}_0$ and 
on the nature of the observable $A$.
The decay constant of the $i$th process, $t_i$, is given by the 
$i$th implied timescale of the transition matrix governing the system's dynamics. 

Calculation of the amplitudes, $\gamma_i$, requires accurate knowledge of the initial state of the system. When this information is not available, the relaxation time can be approximated by the autocorrelation time of $A(t)$, which is given by
\begin{equation}\label{eq:acorr}
\tau_{corr}(A)=\sum_{i=2}^n t_i c_i%
\end{equation}
 where the amplitudes, $c_i$, are closely related with the factors 
$\gamma_i$. See~\cite{noe2011dynamical} for a more detailed derivation.

    \section{Results}

    \subsection{Dinucleotides CC, AC, CA, AA}

\begin{figure} \centering
\centering
    \adjustimage{max size={1.\linewidth}{0.9\paperheight}}{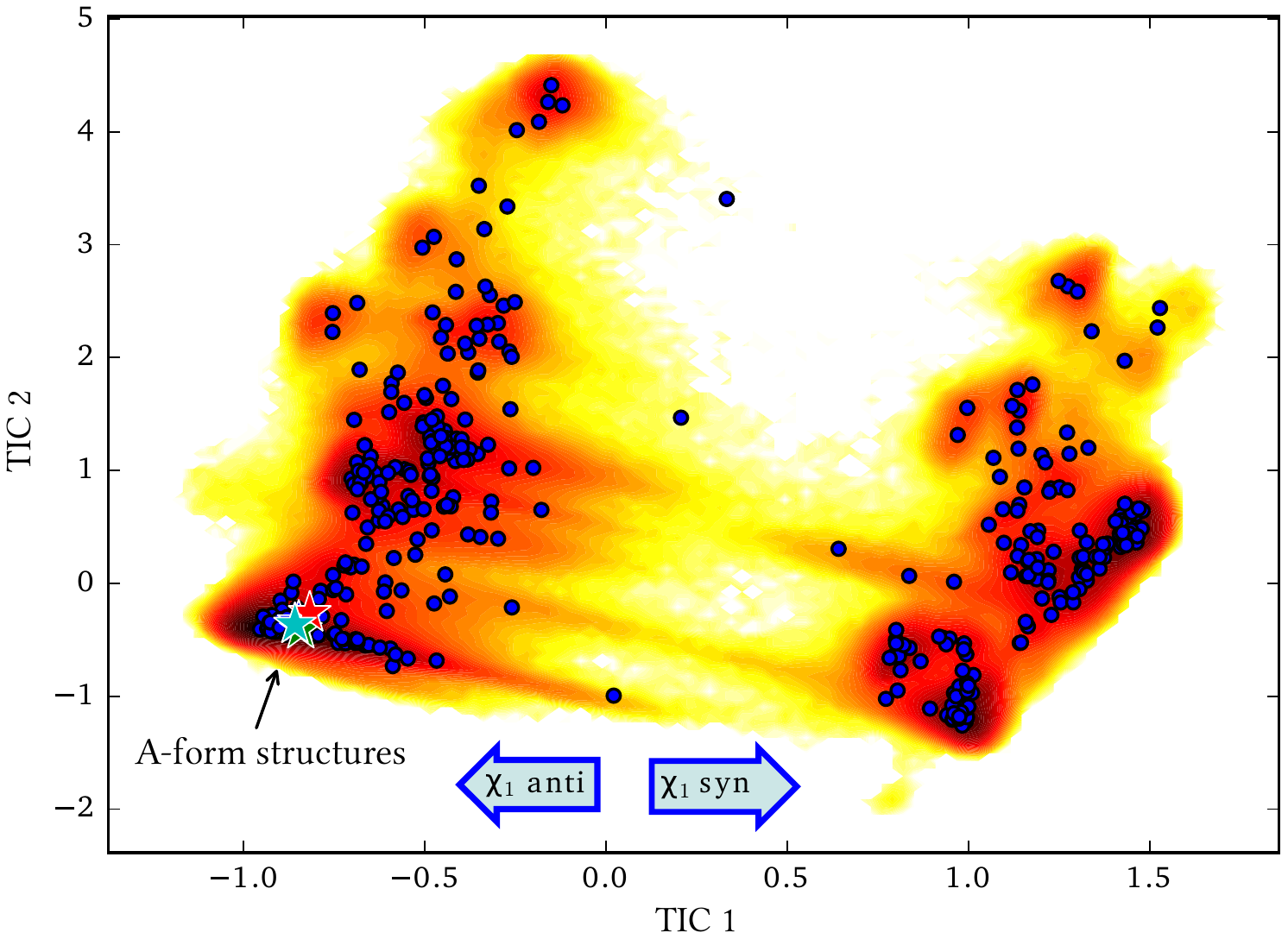}
\caption{2-D histogram of the joint MD data of the 4 dinucleotides, projected on the first 2 TICs; blue circles represent the centers of the microstates obtained from the k-means clustering.
The native A-form structures
are indicated with stars.
}
\label{fig:dino.histo}
\end{figure}

We here report the kinetic analysis performed on all the dinucleotides.
Trajectories for all the investigated dinucleotides were merged together
and analyzed with a single TICA.
TICA provides a low dimensional projection for a complex data set, 
similarly to principal component analysis, but defined so as 
to maximize the autocorrelation times of its components.
The complex phase
space of the different dinucleotides can then be conveniently
projected on the  2-D surface defined by the first two
TICs (see Fig.~\ref{fig:dino.histo}).

    An initial analysis of the TICA components and the trajectories 
shows that the 1st TIC classifies the structures based on the value of
the torsional angle $\chi_1$ relative to the rotation of the glycosidic
bond of the 5' nucleobase (\emph{anti}=negative values,
\emph{syn}=positive values). This suggests that this isomerization 
is the slowest
kinetic process in dinucleotides.

We then constructed a separate MSM for each of the investigated systems.
The convergence of the MSMs was validated by monitoring the convergence of the 
implied timescales  as a function of the lagtime 
(see Fig.~SI~1).
In Tab.~SI~1 we report the slowest timescales of the nine resulting MSMs at the chosen
lag-time of $0.5$~ns.

The four dinucleotides exhibit different timescales. In particular, at temperature $T=277$~K the
largest timescales for CC and CA are in the order of $200$-$300$~ns, whereas
for AA and AC it is around $40$~ns.
The situation is analogous at $T=300$~K, but the timescales are
shorter, in agreement with expectations for higher temperatures.

\begin{figure} \centering
    \adjustimage{max size={0.99\linewidth}{0.6\paperheight}}
{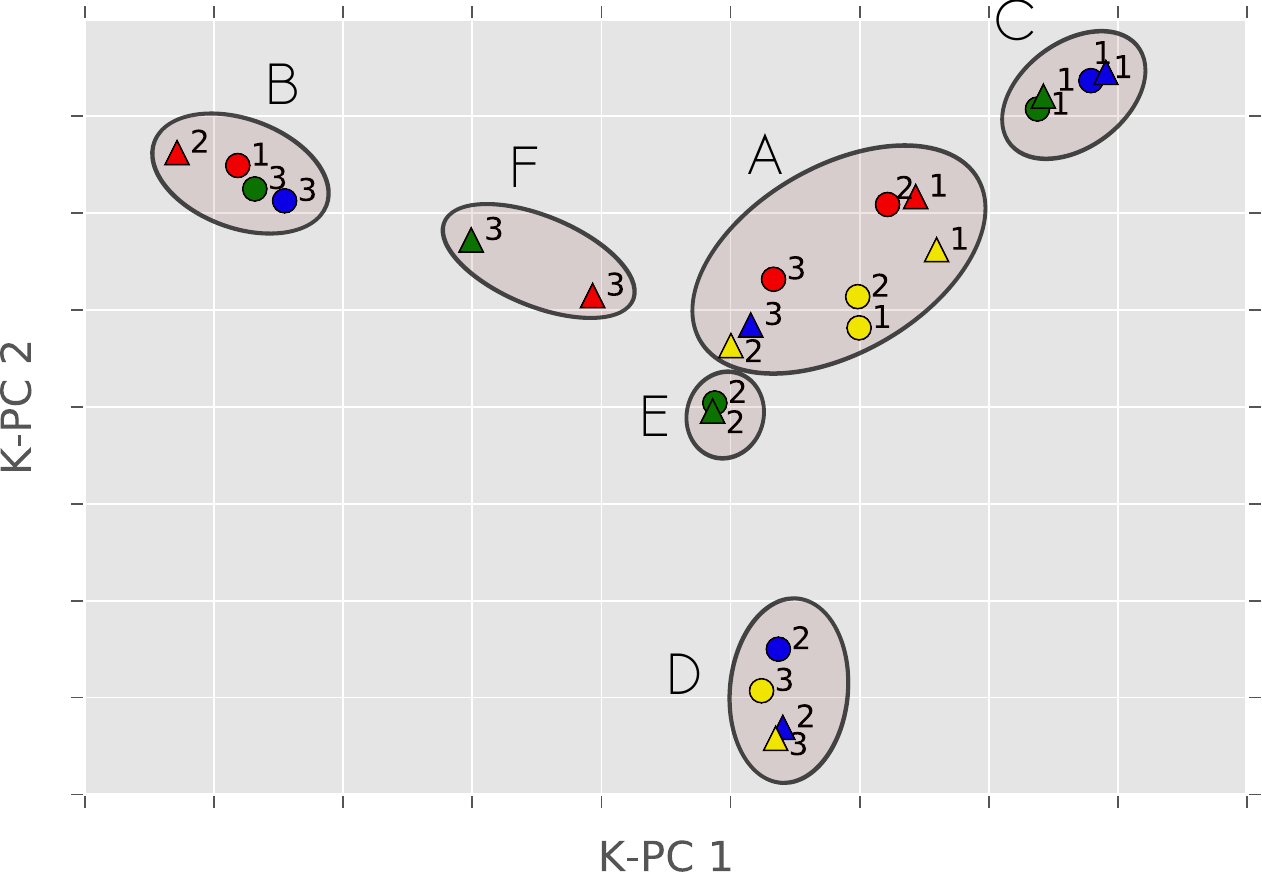}
\caption{First three eigenvectors of each of the eight dinucleotides' MSMs,
projected on the plane
defined by the first two directions identified by KPCA.
Numbers indicate eigenvectors' indexes.
Colors indicate the sequence:
CC (blue); AC (yellow); CA (green); AA (red).
Shapes indicate the simulation conditions:
$T=277$~K (circle), 
$T=300$~K (triangle).
}
\label{fig:dino.kpca}
    \end{figure}

Fig.~SI~2 shows the first three eigenvectors for each of the dinucleotides' MSM,
projected on the first two TICs.
    Since there are a large number of eigenvectors, it is
convenient to exploit the fact that some of them share common features
and define groups of similar processes occurring in different
dinucleotides. In order to do this we evaluate the similarity of two
eigenvectors using the square of their scalar product.
A table summarizing the similarity between the eigenvectors relative to all systems is reported in Fig.~SI~3.
The KPCA algorithm was then used to project them on a 2-D plane where we can easily identify clusters of similar processes
(Fig.~\ref{fig:dino.kpca}).

    Using the information from the 2-D projection shown in Fig.~\ref{fig:dino.kpca} 
and looking at the correlations between each eigenvector and the dihedrals angles (See fig.~SI~4), 
it is possible to identify five groups of eigenvectors that share
similar features between them and are separate from the main group (labeled as $A$)
by the KPCA.
\begin{enumerate}
\item \emph{Group A}
This group collects together all the eigenvectors that are not classified in other groups by the KPCA.

\item \emph{Group B}
These eigenvectors represent the flipping of the
$\chi_1$ torsion. This process is extremely slow ($200$-$300$~ns at $T=277$) when
this nucleobase is a cytosine (CC and CA), while it is much
faster ($<20$~ns) when the base is an adenine (AA and AC).
This effect is likely caused by the stabilization of the \emph{syn}
conformation due to a hydrogen bond between the carbonyl group of
the cytosine at the 5' end and the 5'OH in the corresponding ribose ring.

\item \emph{Group C}
These processes are related to the rotation of the dihedral $\chi_2$.
Cytosines at the 3' end show a much faster dynamics
(\textasciitilde{}$30$-$40$~ns at $T=277$) than those at the 5' end.

\item \emph{Groups D, E, F}
These processes are instead linked to the
formation of specific structures. The conformation of the backbone in
these structures is the same found in RNA Z-helices. 
They are in general
characterized by $\gamma_2$ in $trans$ conformation ($\gamma_2>150$\textdegree\ or
$\gamma_2<-150$\textdegree), and a low distance ($<0.4$ nm) between the
O4' atom on the sugar ring of the 5'-end nucleotide and the center of mass
of the 3' base~\cite{dascenzo2016zdna}.
These three groups  represent the formation 
of Z-motifs, that differ in the orientation of the $\chi_1$,$\chi_2$ glycosidic torsion 
or in the pathway of the process.
\end{enumerate}

We notice that the identification of individual eigenvectors is arbitrary
when the corresponding timescales are comparable within their respective 
statistical errors. 
For a more detailed discussion of this issue see the caption of Fig.~SI~3.

As a further check of the robustness of our results, we  tested the dependence on the ionic concentration of the MSMs of the dinucleotides,
without finding any significative difference (data not shown).

 \subsection{Trinucleotide AAA}

\begin{figure} \centering
          \adjustimage{max size={1.\linewidth}{0.9\paperheight}}{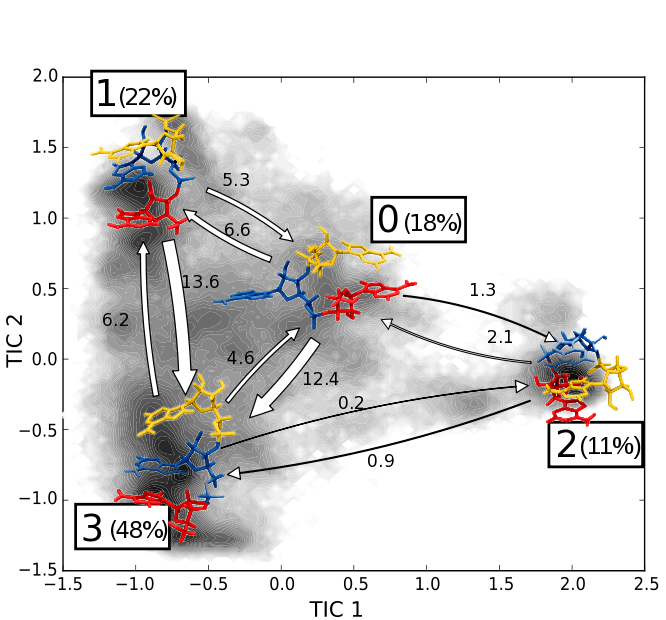}
          \caption{Schematic representation of the 4-state HMM of AAA. A1 (red), A2 (blue), A3 (yellow). 
Percentages indicate the equilibrium population of each state; 
the width of the arrows 
is proportional to the transition rate between the states which are 
also indicated in \textmu s$^{-1}$ units.
Shading indicates the distribution of the simulation data on the TICA plane.
}
    \label{fig:trino.schema}
    \end{figure}

We here report the  MSM obtained for the AAA trinucleotide.
The MSM was validated with the implied timescales test (see Fig.~SI~5).
The MSM of the trinucleotide AAA identified a very slow process
($t=213\pm9$~ns). The fastest processes are dominated by two timescales
around $40$~ns (see Tab.~S2).
    In order to gain further insight on the nature of the first three slow
processes identified by the MSM, we coarse-grained the microstates space
into 4 metastable sets, using an HMM. 
A schematic representation of the HMM is shown in Fig.~\ref{fig:trino.schema} and Fig.~SI~6.

    A first observation about the HMM is that state \#2 corresponds to a
particularly stable state. The transition in and out of this state
has a very large timescale ($200$ to $300$~ns).
The
equilibrium populations of the four states is reported in Fig.~\ref{fig:trino.schema}.

    In order to understand the nature of these four states we analyzed the
distribution of key observables (angles, distances, G-vectors, and stacking score) in the different
HMM states, see Fig.~SI~7, SI~8, SI~9. 
From this analysis we discovered that
state \#2 corresponds to an intercalated structure, in which base A3
stacks between bases A1 and A2.
State \#3 corresponds to the
native state, with a single A-form helix conformation, having all $\chi$ torsions in \emph{anti} conformation. State \#0
acts as an intermediate state, often visited by the system before transitioning
to state \#2. In state \#1 the sequence of stacking interaction is
analogous to state \#3, while the main difference lies in the orientation 
of base A2, that in state \#1  corresponds to a \emph{syn} conformation of the torsion $\chi_2$.

    We also noticed that
a significant fraction of the structures corresponding to state \#0
present an A1-A3 stacking. This state is
nevertheless well separated
kinetically from \#2. In fact, while in state \#2 the stacking of A2-A3
occurs simultaneously with the stacking of A1-A3, in the intermediate
state \#0 the two stacking interactions are never formed together.

We also observed a
recurrent hydrogen bond forming between the non-bridging oxygen of the
phosphate group of base 2 and the 5' hydrogen of base 1, in the intercalated structures. 
Fig.~SI~8 shows the distribution of the distance between these two atoms in the
four metastable states, along with other pairs of atoms that may form hydrogen bonds. 
The formation of this
hydrogen bond is clearly a fundamental step in the formation of
the intercalated structures.

The distribution of the dihedral
angles, particularly the couple
$\alpha_{i+1}$,$\zeta_{i}$, is also informative, as shown in Fig.~SI~9:
\begin{enumerate}
\def\labelenumi{\arabic{enumi}.}
\item
  states \#1 and \#3 are characterized by $\alpha_2$ and $\zeta_1$ $<0$
\item
  state \#2 is characterized by $\alpha_2$ and $\zeta_1$ $>0$
\item
  state \#0 has 
  $\alpha_2$, $\zeta_1$, $\alpha_3$, and $\zeta_2$  $>0$
\end{enumerate}

State \#0 and \#2 are also distinguished by the value of the angle
$\chi_1$ (\emph{syn} in \#0, \emph{high-anti} in \#2). 
The distributions of the three $\gamma$ dihedrals do not vary significantly between the four
metastable states, and we can exclude the presence of kinetically stable Z-motifs,
in contrast with what was observed for the AA dinucleotide.
We also checked that the occupation of any Na$^+$ binding site is lower than 5\% 
in all the metastable states.

\subsection{Tetranucleotide AAAA}

The analysis of the MSM of the AAAA tetranucleotide follows 
a scheme similar to that described for the trinucleotide.
The complexity and the number of available conformations grow exponentially
with the number of bases in an oligonucleotide. For this reason 
it is particularly challenging to sample all the relevant conformational 
space for a tetranucleotide using only plain MD \cite{bergonzo2015highly,Gil-LeyBussi2015}.
In fact, even if our simulations have lengths of several microseconds, many transitions
are observed only once.
This reflects on the quality of the MSM, as it can be seen from the implied timescales plot
(see Fig.~SI~10), and leads to extremely large statistical uncertainties.
From Tab.~\ref{tab:MD} we can also observe that one of the $400$ microstates was not included in the MSM,
since it was visited only once at the beginning of one MD trajectory, so that no entering event was measurable
at the selected lagtime.

Nevertheless it is possible to qualitatively compare 
the predictions of the MSM for AAAA
with those described above for shorter oligonucleotides.

The first two implied timescales exhibited by the system 
(see Tab.~SI~3)
are in the microseconds range 
($3.1\pm 1.1$ and $1.3 \pm 0.6$ \textmu s), and are associated with 
the formation of two different intercalated structures, analogous
to the ones described for the trinucleotide.
For AAAA, 2D NMR spectra show that intercalation is present in less than $5\%$
of the population (see Fig.~SI~11).

\begin{figure} \centering
  \adjustimage{max size={1.\linewidth}{0.9\paperheight}}{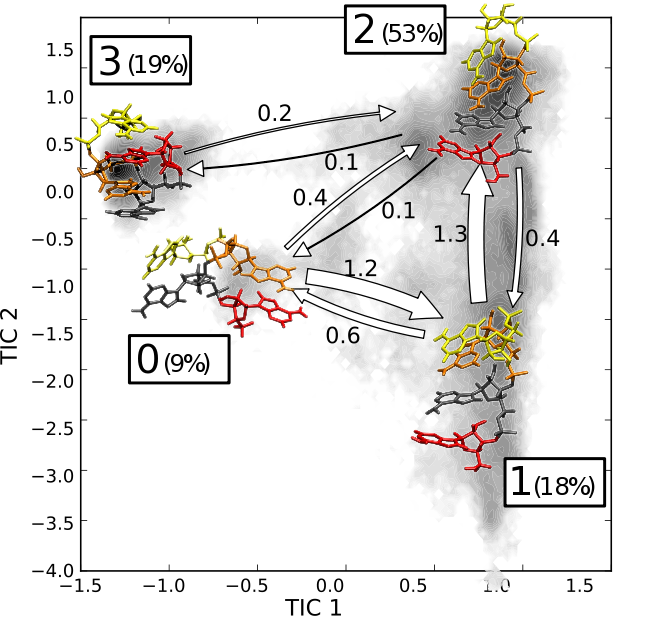}
    \caption{Schematic representation of the 4-state HMM of AAAA. 
A1 (red), A2 (blue), A3 (orange), A4 (yellow).
Percentages indicates the equilibrium population of each state; 
the width of the arrows is proportional to the transition rate between the states which are 
also indicated in \textmu s$^{-1}$ units.
Shading indicates the distribution of the simulation data on the TICA plane.
}
    \label{fig:tetro.schema}
    \end{figure}

Again, to simplify the model we built an HMM,
coarse-graining the MSM into 4 metastable macrostates (see Fig.~SI~12). %
Fig.~\ref{fig:tetro.schema} shows a schematic representation of the HMM projected on the
first two TICs.
Also in this case the TICA identifies 
the formation of the intercalated structures (State \#3)
as the slowest process.

Two of the resulting states 
(\#1 and \#2) display a canonical stacking pattern whereas the other two states
(\#3 and \#0) are characterized by the stacking of non-consecutive bases (see Fig.~SI~13). 
Specifically, state \#2 contains the canonical A-form helix, whereas in state \#1 base A3 is 	flipped to \emph{syn} conformation.
States \#3 and \#0 instead are distinguished by their stacking pattern,
and they share the same features reported for trinucleotides, that is, 
$\alpha$ and $\zeta$ in $g+$ conformation and 
the presence of stabilizing hydrogen bonds with non-bridging oxygens (Fig.~SI~14, SI~15).
State \#3 contains intercalated structures 
analogous to the one reported in previous 
works~\cite{yildirim2011benchmarking,bergonzo2015highly,condon2015stacking,gil2016empirical,BottaroGil-LeyBussi2016}, while state \#0 presents
A2 and A4 flipped out and stacked on each other. %
It is reasonable to expect other combinations of base orientation
and stackings to arise
when increasing the sampling.

Unfortunately the large statistical errors in the HMM timescales make it difficult 
to discriminate quantitatively the different processes, and to clearly assign an implied timescale to 
each of them.%

\subsection{Comparison with Temperature-jump experiments}
The timescales predicted by our MSMs 
can be compared with relaxation times 
measured using T-jump experiments in~\cite{porschke1978molecular}.
A proper comparison should follow the procedure
 explained in~\cite{noe2011dynamical},
where the relaxation of an experimental observable can be decomposed in exponential contributions coming from each MSM eigenvector.
This requires knowledge of the experimental observable.
In~\cite{porschke1978molecular} the relaxation is measured with
UV absorption. We modeled this using the stacking score
proposed in~\cite{condon2015stacking}.
To estimate the relaxation rate without the need of further
assumptions on the equilibrium distribution of the systems 
prior to T-jump,
we computed the autocorrelation time of the stacking score. %

The results of this calculation are reported in Tab.~\ref{tab:tau}, 
along with the experimental relaxation times measured in~\cite{porschke1978molecular}.
For AAA and AAAA we report the
autocorrelation time (Eq.~3) relative to the slowest stacking interaction
(A1-A2 for AAA, A1-A3 for AAAA). 
We excluded from the calculations the contributions %
of the formation of intercalated/non-canonical structures,
since the increased stability of such structures is a known limitation
of the current force field~\cite{yildirim2011benchmarking,bergonzo2015highly,condon2015stacking,gil2016empirical,BottaroGil-LeyBussi2016}.
This was done by using Eq.~3\ and setting to zero the amplitudes, $\gamma_i$, of the relative processes.
As a cross check for this procedure we also constructed a new MSM where
we remove from the active set the intercalated microstates,
identified from the analysis of the sign
of the relevant eigenvectors of the original MSM.
The resulting timescales and autocorrelation times are within the statistical error
of the ones derived with the original MSM (data not shown).
\begin{table}
\centering
    \begin{tabular}{lcc}
Molecule & $\tau$ (ns) & Exp. (ns)~\cite{porschke1978molecular} \\
\hline
\multicolumn{2}{l}{$T=277$~K}\\
\hline
CC & $24  \pm 2$  & $30 \pm 6$\\
AC & $9   \pm 1$  & $42 \pm 8$\\
CA & $25  \pm 1$  & $30 \pm 6$\\
AA & $9.1 \pm 0.2$& \:\:$50 \pm 10$\\
AAAA & $171^\textrm{a,b} \pm 26$ & - \\
\hline
\multicolumn{2}{l}{$T=300$~K}\\
\hline
CC & $7.2 \pm 0.4$ & - \\
AC & $3.44 \pm 0.03$ & - \\
CA & $4.8 \pm 0.1$ & - \\
AA & $3.36 \pm 0.04$ & $29^\textrm{c} \pm 6$\\
AAA & $25.8 \pm 0.4$ & $45^\textrm{c} \pm 9$\\
\multirow{2}{*}{AAAA} & - & $270^\textrm{c,d} \pm 27$ ($20\%$)\\
                      & - & $47^\textrm{c,d} \pm 5$   ($80\%$)\\
\hline
\end{tabular}
\caption{Autocorrelation times of the stacking score predicted by
the different MSMs, compared with the experimental relaxation times 
($1$~M Na$^+$).
$^\textrm{a}$ Simulations performed at $275$~K.
$^\textrm{b}$ Value for A1-A3 a stack not observed by 
NMR~\cite{condon2015stacking};  values for A1-A2, A2-A3, and A3-A4 are 
$78 \pm 16$, $81 \pm 16$, and $92 \pm 26$~ns, respectively.
$^\textrm{c}$ Experiments performed at $297$~K.
$^\textrm{d}$ Ref.~\cite{porschke1978molecular} reports two relaxation
times, with the relative amplitudes shown in brackets.
}
\label{tab:tau}
\end{table}

We notice that the contributions of the slow modes of CC and CA to the
stacking score kinetics are extremely small, since the autocorrelation
time is almost ten times shorter than the associated timescale (Tab.~SI~1). 
The values of $\tau_{corr}$ predicted for CC, CA, AAA and AAAA are 
in good agreement with the experimental relaxation times.
On the other hand, the values obtained for AA and AC are significantly
shorter than the experimental values.

The T-jump relaxations at $297$~K reported in~\cite{porschke1978molecular} 
also included a long relaxation time of $600$-$900$~ns 
for A$_2$, A$_3$, A$_4$, A$_5$, and A$_{14}$ in $1$~M Na$^+$ when the transition was probed with $> 280$~nm light.  
In the same study, relaxation times of $200 \pm 20$~ns and $700 \pm 140$~ns 
were reported for poly(A) in $0.2$~M Na$^+$.  
These experiments were conducted with a cable discharge temperature jump apparatus 
where up to $200$ kV/cm is transiently applied to the sample~\cite{porschke1976cable}.
An independent study using a laser induced temperature jump of poly(A) in $0.2$~M Na$^+$, $T=298$~K, 
however, reported only a $270 \pm 70$~ns relaxation at $285$~nm~\cite{dewey1979laser}.
None of the MD simulations of A$_2$ and A$_3$
generated a timescale longer than $300$~ns.
It is therefore possible that the high electric field 
in the cable discharge experiments somehow affected the RNA,
leading to the appearance of an artifactual relaxation process.

    \section{Discussion}

    What can we learn from this analysis about the kinetic properties of
oligonucleotides?

    The slow implied timescales observed for CC and CA are one order 
of magnitude longer than the experimental relaxation times (Tab.~SI~1).
These timescales are related to the transition from \emph{anti} to
\emph{syn} of the cytosine at the 5' end and are likely caused by the formation
of a hydrogen bond between the carbonyl group of
the cytosine at the 5' end and the 5'OH in the corresponding ribose ring.
An explanation for this inconsistency may
be found in the inaccuracy of the force field, which is a known
limitation in the field of MD simulations of RNA.
\emph{Syn} cytosines are rare in non-catalytic RNAs~\cite{sokoloski2011prevalence} and may be over represented in simulations.
However, we notice that it was not necessary to remove these timescales
to obtain a reasonable agreement between the autocorrelation time of the
stacking score and the experimental relaxation times. This
is essentially due to the fact that these slow \emph{syn}/\emph{anti}
flips are poorly correlated with the stacking score.

The slowest  timescales observed in AAA and AAAA are related to the
formation of kinetically stable intercalated
structures.
For AAAA the presence of a significant population of intercalated structures in solution
has been ruled out by careful interpretation of the NOE data~\cite{yildirim2011benchmarking,condon2015stacking}.
In these experiments, all the observed signals have similar linewidths
and only non-exchangeable protons are analyzed allowing the contacts predicted from
the intercalated structures and not observed in the experimental spectrum to be used
in the characterization of the structural ensemble~\cite{zagrovic2006comparing}.
A portion of the 2D NOESY spectrum of AAAA is reported in Fig.~SI~11.
This inconsistency
suggests that this metastable structure is an artifact of the
simulation, likely caused by an imperfect parametrization of the force
field.
Analysis of the intercalated states
revealed some structural details that seem to play a crucial role in
stabilizing these structures. In particular these are 1) the
formation of a hydrogen bond between the non-bridging oxygen of the
phosphate group of one nucleotide and a hydroxyl from another nucleotide~\cite{condon2015stacking}
, 2) the transition from negative to positive of the
torsional angles $\alpha_{i+1}$,$\zeta_{i}$ (see Fig.~SI~9, SI~15).
We propose that this information must be kept in mind when trying to
modify the parameters to improve force-field accuracy. In particular
it has been shown that tuning the parametrization of the dihedrals $\alpha$
and $\zeta$ significantly improves agreement with NOE data for several tetranucleotides and tetraloops~\cite{gil2016empirical,bottaro2016free,cesari2016combining}.
We observe that the overstabilization of intercalated structures might also
be related to an overestimation of stacking interactions in the AMBER force field that
has been suggested in ref.~\cite{chen2013high}.
In the present work, to obtain reasonable kinetic properties
without performing new simulations with a corrected force field
we found it necessary to remove
these structures from the ensemble.

Once the unphysical structures and transitions have been removed from the MSM,
it is possible to use the remaining eigenvalues and eigenvectors to
estimate the experimental relaxation times.
It is important to recall that for an appropriate comparison
with experimental data it is necessary to define an observable
that is proportional to the measured intensity.
We here used the stacking score defined in ref.~\cite{condon2015stacking}.
The autocorrelation time of this score is reported in Tab.~\ref{tab:tau}
and
can be directly compared with experiments,
with which it is in good agreement.
It must be mentioned that the self-diffusion coefficient of the TIP3P
water model is approximately 2.5 times larger than the experimental
one~\cite{mills1973self,mark2001structure}. This might induce an artificial acceleration of processes
depending on water rearrangement.
Considering experimental error, the largest difference
between autocorrelation times and experimental relaxation rates
may be as small as 7-fold, corresponding to a difference 
of $1$ kcal/mol in activation free energy at $300$~K.

The virtual removal of structures that we considered to be artificially stabilized
by inaccuracies in the force field was instrumental in achieving an agreement with experimental
timescales. 
This removal has been justified by comparison with NMR experiments.
We suggest that force field refinements should be mainly driven
by comparison with equilibrium experiments in solution,
as it has been done in a very recent work \cite{cesari2016combining}.
Hypothetically, a correction to the force field
that fixes these problems might also affect the transition rates. However,
corrections designed to penalize specific rotamers as those used in
refs \cite{gil2016empirical,bottaro2016free,cesari2016combining} are not expected
to affect the transition rates between non-penalized  metastable states.
Moreover, once the force field has been refined so as to provide
equilibrium populations in agreement with solution data, possible mismatches
between the predicted time scales and the relaxation times observed in
experiments might be used to identify errors in the parametrization
of energetic barriers and suggest further refinements.

For the dinucleotides and trinucleotide we additionally assessed the influence of
Na$^+$ ions.
Our results confirm that
monovalent ions do not play any significant role in the kinetic of RNA oligonucleotides, as was also reported in Ref.~\cite{porschke1978molecular}.
We notice that in a previous work we did not find a significant dependence of RNA dynamics on monovalent ion concentration \cite{pinamonti2015elastic}.

In general,
the predicted relaxation time in all the considered systems
 is not determined by the rate of the
helix $\leftrightarrow$ coil
transition.
It is instead related to the rate of transitions between different
structures, stabilized by stacking or other kinds of interactions.
Examples of this are the Z-motifs in dinucleotides, or
helices with flipped nucleotides in longer sequences.
This suggests that the timescales obtained from relaxation experiments of
oligonucleotides 
may be due to transitions between different ``folded states'', rather than between stacked, native structure and random coil. That is, these relaxation rates are dominated by the
presence of various ``kinetic traps'', i.e.~kinetically stable
structures where the system may get stuck for a relatively long time
before being able to reach its minimum free-energy conformation.

This paper demonstrates how MD simulations and MSMs 
can be used to provide deeper interpretation of experimental 
measurements of the kinetics of RNA folding.  
While most experimental methods report weighted averages 
for an ensemble of structures, MD follows transitions 
of a single molecule at atomistic resolution.
If the simulations are statistically converged, then the results from MSM analysis 
can be compared to experimental measurements.  
Although current RNA force fields do not accurately predict 
structural ensembles~\cite{bergonzo2015highly,condon2015stacking}, 
the results presented here suggest that the latest AMBER force field 
can reproduce the order of magnitude of T-jump relaxation times~\cite{dewey1979laser,porschke1978molecular} 
measured for short oligonucleotides.  
Most current tests of force fields use structural data from NMR and x-ray diffraction as benchmarks.  The results presented above indicate 
that comparison to experimental kinetic data can provide 
new benchmarks in the future.  
Moreover, when MD simulations accurately predict structures, 
they can generate more detailed interpretations of experiments 
and suggest new experiments to test hypotheses.

\section{Other Information Available}
Trajectories and analysis scripts used in this paper are available at \url{http://github.com/srnas/oligonucleotides-kinetics}.

\section{Acknowledgement}
G.P. and G.B. received support by the European Research Council -- Starting Grant 306662, S-RNA-S.
D.T., D.C., and J.Z. were supported by the NIH grant GM22939; 
F.P. and F.N. have received funding from the
European Research Council -- Starting Grant ``pcCell''.
F.P. also acknowledges funding from Grant
Deutsche Forschungsgemeinschaft SFB1114.
We are grateful to Dr. Sandro Bottaro for many helpful and 
stimulating conversations, and to Dr. Scott Kennedy for help
interpreting NMR of AAAA.
We thank all the members of No\'e group for discussion and advice.

\begin{suppinfo}
Three tables reporting the values of the timescales of the slowest processes identified by the MSMs for all the systems (Tab.~SI~1-3).
Supporting figures for the dinucleotides' kinetic analysis: convergence of the MSM implied timescales (Fig. SI~1),
graphical representation of the eigenvectors (Fig.~SI~2), similarity between the first three eigenvectors of each system (Fig.~SI~3), 
correlation between the first eigenvectors and the torsional angles (Fig.~SI~4).
Supporting figures for the AAA trinucleotide kinetic analysis: convergence of the MSM implied timescales (Fig.~SI~5),
representative structures of the HMM states (Fig.~SI~6),
stacking patterns (Fig.~SI~7), hydrogen bond formation (Fig.~SI~8), and torsional angles distribution (Fig.~SI~9).
One figure reporting a portion of 2D NOESY spectrum of AAAA (Fig.~SI~11).
Supporting figures for the AAAA tetranucleotide kinetic analysis: convergence of the MSM implied timescales (Fig.~SI~10),
representative structures of the HMM states (Fig.~SI~12),
stacking patterns (Fig.~SI~13), hydrogen bond formation (Fig.~SI~14), and torsional angles distribution (Fig.~SI~15).

\end{suppinfo}
\providecommand*\mcitethebibliography{\thebibliography}
\csname @ifundefined\endcsname{endmcitethebibliography}
  {\let\endmcitethebibliography\endthebibliography}{}


\begin{mcitethebibliography}{62}
\providecommand*\natexlab[1]{#1}
\providecommand*\mciteSetBstSublistMode[1]{}
\providecommand*\mciteSetBstMaxWidthForm[2]{}
\providecommand*\mciteBstWouldAddEndPuncttrue
  {\def\EndOfBibitem{\unskip.}}
\providecommand*\mciteBstWouldAddEndPunctfalse
  {\let\EndOfBibitem\relax}
\providecommand*\mciteSetBstMidEndSepPunct[3]{}
\providecommand*\mciteSetBstSublistLabelBeginEnd[3]{}
\providecommand*\EndOfBibitem{}
\mciteSetBstSublistMode{f}
\mciteSetBstMaxWidthForm{subitem}{(\alph{mcitesubitemcount})}
\mciteSetBstSublistLabelBeginEnd
  {\mcitemaxwidthsubitemform\space}
  {\relax}
  {\relax}

\bibitem[Morris and Mattick(2014)Morris, and Mattick]{morris2014riseregRNA}
Morris,~K.~V.; Mattick,~J.~S. \emph{Nat. Rev. Genet.} \textbf{2014}, \emph{15},
  423--437\relax
\mciteBstWouldAddEndPuncttrue
\mciteSetBstMidEndSepPunct{\mcitedefaultmidpunct}
{\mcitedefaultendpunct}{\mcitedefaultseppunct}\relax
\EndOfBibitem
\bibitem[Al-Hashimi and Walter(2008)Al-Hashimi, and Walter]{al2008rna}
Al-Hashimi,~H.~M.; Walter,~N.~G. \emph{Curr. Opin. Struct. Biol.}
  \textbf{2008}, \emph{18}, 321--329\relax
\mciteBstWouldAddEndPuncttrue
\mciteSetBstMidEndSepPunct{\mcitedefaultmidpunct}
{\mcitedefaultendpunct}{\mcitedefaultseppunct}\relax
\EndOfBibitem
\bibitem[Bloomfield et~al.(2000)Bloomfield, Crothers, and
  Tinoco]{bloomfield2000nucleic}
Bloomfield,~V.; Crothers,~D.; Tinoco,~I. \emph{Nucleic Acids: Structures,
  Properties, and Functions}; University Science Books, 2000\relax
\mciteBstWouldAddEndPuncttrue
\mciteSetBstMidEndSepPunct{\mcitedefaultmidpunct}
{\mcitedefaultendpunct}{\mcitedefaultseppunct}\relax
\EndOfBibitem
\bibitem[Hobza and {\v{S}}poner(1999)Hobza, and
  {\v{S}}poner]{hobza1999structure}
Hobza,~P.; {\v{S}}poner,~J. \emph{Chem. Rev.} \textbf{1999}, \emph{99},
  3247--3276\relax
\mciteBstWouldAddEndPuncttrue
\mciteSetBstMidEndSepPunct{\mcitedefaultmidpunct}
{\mcitedefaultendpunct}{\mcitedefaultseppunct}\relax
\EndOfBibitem
\bibitem[Lee et~al.(1976)Lee, Ezra, Kondo, Sarma, and
  Danyluk]{lee1976conformational}
Lee,~C.-H.; Ezra,~F.~S.; Kondo,~N.~S.; Sarma,~R.~H.; Danyluk,~S.~S.
  \emph{Biochemistry} \textbf{1976}, \emph{15}, 3627--3639\relax
\mciteBstWouldAddEndPuncttrue
\mciteSetBstMidEndSepPunct{\mcitedefaultmidpunct}
{\mcitedefaultendpunct}{\mcitedefaultseppunct}\relax
\EndOfBibitem
\bibitem[Ezra et~al.(1977)Ezra, Lee, Kondo, Danyluk, and
  Sarma]{ezra1977conformational}
Ezra,~F.~S.; Lee,~C.-H.; Kondo,~N.~S.; Danyluk,~S.~S.; Sarma,~R.~H.
  \emph{Biochemistry} \textbf{1977}, \emph{16}, 1977--1987\relax
\mciteBstWouldAddEndPuncttrue
\mciteSetBstMidEndSepPunct{\mcitedefaultmidpunct}
{\mcitedefaultendpunct}{\mcitedefaultseppunct}\relax
\EndOfBibitem
\bibitem[Lee and Tinoco(1980)Lee, and Tinoco]{lee1980conformation}
Lee,~C.-H.; Tinoco,~I. \emph{Biophys. Chem.} \textbf{1980}, \emph{11},
  283--294\relax
\mciteBstWouldAddEndPuncttrue
\mciteSetBstMidEndSepPunct{\mcitedefaultmidpunct}
{\mcitedefaultendpunct}{\mcitedefaultseppunct}\relax
\EndOfBibitem
\bibitem[Olsthoorn et~al.(1982)Olsthoorn, Doornbos, Leeuw, and
  Altona]{olsthoorn1982influence}
Olsthoorn,~C.~S.; Doornbos,~J.; Leeuw,~H.~P.; Altona,~C. \emph{Eur. J.
  Biochem.} \textbf{1982}, \emph{125}, 367--382\relax
\mciteBstWouldAddEndPuncttrue
\mciteSetBstMidEndSepPunct{\mcitedefaultmidpunct}
{\mcitedefaultendpunct}{\mcitedefaultseppunct}\relax
\EndOfBibitem
\bibitem[Lee(1983)]{lee1983conformational}
Lee,~C.-H. \emph{Eur. J. Biochem.} \textbf{1983}, \emph{137}, 347--356\relax
\mciteBstWouldAddEndPuncttrue
\mciteSetBstMidEndSepPunct{\mcitedefaultmidpunct}
{\mcitedefaultendpunct}{\mcitedefaultseppunct}\relax
\EndOfBibitem
\bibitem[Vokacova et~al.(2009)Vokacova, Budes{\'\i}nsk{\'y}, Rosenberg,
  Schneider, {\v{S}}poner, and Sychrovsky]{vokacova2009structure}
Vokacova,~Z.; Budes{\'\i}nsk{\'y},~M.; Rosenberg,~I.; Schneider,~B.;
  {\v{S}}poner,~J.; Sychrovsky,~V. \emph{J. Phys. Chem. B} \textbf{2009},
  \emph{113}, 1182--1191\relax
\mciteBstWouldAddEndPuncttrue
\mciteSetBstMidEndSepPunct{\mcitedefaultmidpunct}
{\mcitedefaultendpunct}{\mcitedefaultseppunct}\relax
\EndOfBibitem
\bibitem[Yildirim et~al.(2011)Yildirim, Stern, Tubbs, Kennedy, and
  Turner]{yildirim2011benchmarking}
Yildirim,~I.; Stern,~H.~A.; Tubbs,~J.~D.; Kennedy,~S.~D.; Turner,~D.~H.
  \emph{J. Phys. Chem. B} \textbf{2011}, \emph{115}, 9261--9270\relax
\mciteBstWouldAddEndPuncttrue
\mciteSetBstMidEndSepPunct{\mcitedefaultmidpunct}
{\mcitedefaultendpunct}{\mcitedefaultseppunct}\relax
\EndOfBibitem
\bibitem[Tubbs et~al.(2013)Tubbs, Condon, Kennedy, Hauser, Bevilacqua, and
  Turner]{tubbs2013nuclear}
Tubbs,~J.~D.; Condon,~D.~E.; Kennedy,~S.~D.; Hauser,~M.; Bevilacqua,~P.~C.;
  Turner,~D.~H. \emph{Biochemistry} \textbf{2013}, \emph{52}, 996--1010\relax
\mciteBstWouldAddEndPuncttrue
\mciteSetBstMidEndSepPunct{\mcitedefaultmidpunct}
{\mcitedefaultendpunct}{\mcitedefaultseppunct}\relax
\EndOfBibitem
\bibitem[Condon et~al.(2015)Condon, Kennedy, Mort, Kierzek, Yildirim, and
  Turner]{condon2015stacking}
Condon,~D.~E.; Kennedy,~S.~D.; Mort,~B.~C.; Kierzek,~R.; Yildirim,~I.;
  Turner,~D.~H. \emph{J. Chem. Theory Comput.} \textbf{2015}, \emph{11},
  2729--2742\relax
\mciteBstWouldAddEndPuncttrue
\mciteSetBstMidEndSepPunct{\mcitedefaultmidpunct}
{\mcitedefaultendpunct}{\mcitedefaultseppunct}\relax
\EndOfBibitem
\bibitem[P{\"o}rschke(1976)]{porschke1976cable}
P{\"o}rschke,~D. \emph{Rev. Sci. Instrum.} \textbf{1976}, \emph{47},
  1363--1365\relax
\mciteBstWouldAddEndPuncttrue
\mciteSetBstMidEndSepPunct{\mcitedefaultmidpunct}
{\mcitedefaultendpunct}{\mcitedefaultseppunct}\relax
\EndOfBibitem
\bibitem[P{\"o}rschke(1978)]{porschke1978molecular}
P{\"o}rschke,~D. \emph{Biopolymers} \textbf{1978}, \emph{17}, 315--323\relax
\mciteBstWouldAddEndPuncttrue
\mciteSetBstMidEndSepPunct{\mcitedefaultmidpunct}
{\mcitedefaultendpunct}{\mcitedefaultseppunct}\relax
\EndOfBibitem
\bibitem[Dewey and Turner(1979)Dewey, and Turner]{dewey1979laser}
Dewey,~T.; Turner,~D.~H. \emph{Biochemistry} \textbf{1979}, \emph{18},
  5757--5762\relax
\mciteBstWouldAddEndPuncttrue
\mciteSetBstMidEndSepPunct{\mcitedefaultmidpunct}
{\mcitedefaultendpunct}{\mcitedefaultseppunct}\relax
\EndOfBibitem
\bibitem[{\v{S}}poner et~al.(2014){\v{S}}poner, Ban{\'a}{\v{s}}, Jure{\v{c}}ka,
  Zgarbov{\'a}, K{\"u}hrov{\'a}, Havrila, Krepl, Stadlbauer, and
  Otyepka]{sponer2014molecular}
{\v{S}}poner,~J.; Ban{\'a}{\v{s}},~P.; Jure{\v{c}}ka,~P.; Zgarbov{\'a},~M.;
  K{\"u}hrov{\'a},~P.; Havrila,~M.; Krepl,~M.; Stadlbauer,~P.; Otyepka,~M.
  \emph{J. Phys. Chem. Lett.} \textbf{2014}, \emph{5}, 1771--1782\relax
\mciteBstWouldAddEndPuncttrue
\mciteSetBstMidEndSepPunct{\mcitedefaultmidpunct}
{\mcitedefaultendpunct}{\mcitedefaultseppunct}\relax
\EndOfBibitem
\bibitem[Bergonzo et~al.(2013)Bergonzo, Henriksen, Roe, Swails, Roitberg, and
  Cheatham~III]{bergonzo2013multidimensional}
Bergonzo,~C.; Henriksen,~N.~M.; Roe,~D.~R.; Swails,~J.~M.; Roitberg,~A.~E.;
  Cheatham~III,~T.~E. \emph{J. Chem. Theory Comput.} \textbf{2013}, \emph{10},
  492--499\relax
\mciteBstWouldAddEndPuncttrue
\mciteSetBstMidEndSepPunct{\mcitedefaultmidpunct}
{\mcitedefaultendpunct}{\mcitedefaultseppunct}\relax
\EndOfBibitem
\bibitem[Bergonzo et~al.(2015)Bergonzo, Henriksen, Roe, and
  Cheatham]{bergonzo2015highly}
Bergonzo,~C.; Henriksen,~N.~M.; Roe,~D.~R.; Cheatham,~T.~E. \emph{RNA}
  \textbf{2015}, \emph{21}, 1578--1590\relax
\mciteBstWouldAddEndPuncttrue
\mciteSetBstMidEndSepPunct{\mcitedefaultmidpunct}
{\mcitedefaultendpunct}{\mcitedefaultseppunct}\relax
\EndOfBibitem
\bibitem[Gil-Ley and Bussi(2015)Gil-Ley, and Bussi]{Gil-LeyBussi2015}
Gil-Ley,~A.; Bussi,~G. \emph{J. Chem. Theory Comput.} \textbf{2015}, \emph{11},
  1077--1085\relax
\mciteBstWouldAddEndPuncttrue
\mciteSetBstMidEndSepPunct{\mcitedefaultmidpunct}
{\mcitedefaultendpunct}{\mcitedefaultseppunct}\relax
\EndOfBibitem
\bibitem[Brown et~al.(2015)Brown, Andrews, and Elcock]{brown2015stacking}
Brown,~R.~F.; Andrews,~C.~T.; Elcock,~A.~H. \emph{J. Chem. Theory Comput.}
  \textbf{2015}, \emph{11}, 2315--2328\relax
\mciteBstWouldAddEndPuncttrue
\mciteSetBstMidEndSepPunct{\mcitedefaultmidpunct}
{\mcitedefaultendpunct}{\mcitedefaultseppunct}\relax
\EndOfBibitem
\bibitem[Gil-Ley et~al.(2016)Gil-Ley, Bottaro, and Bussi]{gil2016empirical}
Gil-Ley,~A.; Bottaro,~S.; Bussi,~G. \emph{J. Chem. Theory Comput.}
  \textbf{2016}, \emph{12}, 2790--2798\relax
\mciteBstWouldAddEndPuncttrue
\mciteSetBstMidEndSepPunct{\mcitedefaultmidpunct}
{\mcitedefaultendpunct}{\mcitedefaultseppunct}\relax
\EndOfBibitem
\bibitem[Cesari et~al.(2016)Cesari, Gil-Ley, and Bussi]{cesari2016combining}
Cesari,~A.; Gil-Ley,~A.; Bussi,~G. \emph{J. Chem. Theory Comput.}
  \textbf{2016}, \emph{12}, 6192--6200\relax
\mciteBstWouldAddEndPuncttrue
\mciteSetBstMidEndSepPunct{\mcitedefaultmidpunct}
{\mcitedefaultendpunct}{\mcitedefaultseppunct}\relax
\EndOfBibitem
\bibitem[Colizzi and Bussi(2012)Colizzi, and Bussi]{colizzi2012rna}
Colizzi,~F.; Bussi,~G. \emph{J. Am. Chem. Soc.} \textbf{2012}, \emph{134},
  5173--5179\relax
\mciteBstWouldAddEndPuncttrue
\mciteSetBstMidEndSepPunct{\mcitedefaultmidpunct}
{\mcitedefaultendpunct}{\mcitedefaultseppunct}\relax
\EndOfBibitem
\bibitem[H{\"a}se and Zacharias(2016)H{\"a}se, and Zacharias]{hase2016free}
H{\"a}se,~F.; Zacharias,~M. \emph{Nucleic Acids Res.} \textbf{2016}, \emph{44},
  7100--108\relax
\mciteBstWouldAddEndPuncttrue
\mciteSetBstMidEndSepPunct{\mcitedefaultmidpunct}
{\mcitedefaultendpunct}{\mcitedefaultseppunct}\relax
\EndOfBibitem
\bibitem[Bowman et~al.(2008)Bowman, Huang, Yao, Sun, Carlsson, Guibas, and
  Pande]{bowman2008structural}
Bowman,~G.~R.; Huang,~X.; Yao,~Y.; Sun,~J.; Carlsson,~G.; Guibas,~L.~J.;
  Pande,~V.~S. \emph{J. Am. Chem. Soc.} \textbf{2008}, \emph{130},
  9676--9678\relax
\mciteBstWouldAddEndPuncttrue
\mciteSetBstMidEndSepPunct{\mcitedefaultmidpunct}
{\mcitedefaultendpunct}{\mcitedefaultseppunct}\relax
\EndOfBibitem
\bibitem[Pronk et~al.(2013)Pronk, P{\'a}ll, Schulz, Larsson, Bjelkmar,
  Apostolov, Shirts, Smith, Kasson, van~der Spoel, Hess, and
  Lindhal]{pronk2013gromacs}
Pronk,~S.; P{\'a}ll,~S.; Schulz,~R.; Larsson,~P.; Bjelkmar,~P.; Apostolov,~R.;
  Shirts,~M.~R.; Smith,~J.~C.; Kasson,~P.~M.; van~der Spoel,~D.; Hess,~B.;
  Lindhal,~E. \emph{Bioinformatics} \textbf{2013}, \emph{29}, 845--854\relax
\mciteBstWouldAddEndPuncttrue
\mciteSetBstMidEndSepPunct{\mcitedefaultmidpunct}
{\mcitedefaultendpunct}{\mcitedefaultseppunct}\relax
\EndOfBibitem
\bibitem[Case et~al.(2011)Case, Darden, Cheatham, Simmerling, Wang, Duke, Luo,
  Crowley, Walker, Zhang, Merz, Wang, Hayik, Roitberg, Seabra, Kolossv\'{a}ry,
  Wong, Paesani, Vanicek, Wu, Brozell, Steinbrecher, Gohlke, Yang, Tan, Mongan,
  Hornak, Cui, Mathews, Seetin, Sagui, Babin, and Kollman]{case2011amber}
Case,~D.~A.; Darden,~T.~A.; Cheatham,~T.~E.; Simmerling,~C.~L.; Wang,~J.;
  Duke,~R.~E.; Luo,~R.; Crowley,~M.; Walker,~R.~C.; Zhang,~W.; Merz,~K.~M.;
  Wang,~B.; Hayik,~S.; Roitberg,~A.; Seabra,~G.; Kolossv\'{a}ry,~I.;
  Wong,~K.~F.; Paesani,~F.; Vanicek,~J.; Wu,~X.; Brozell,~S.~R.;
  Steinbrecher,~T.; Gohlke,~H.; Yang,~L.; Tan,~C.; Mongan,~J.; Hornak,~V.;
  Cui,~G.; Mathews,~D.~H.; Seetin,~M.~G.; Sagui,~C.; Babin,~V.; Kollman,~P.~A.
  {AMBER 11}. 2011; \url{http://ambermd.org/}\relax
\mciteBstWouldAddEndPuncttrue
\mciteSetBstMidEndSepPunct{\mcitedefaultmidpunct}
{\mcitedefaultendpunct}{\mcitedefaultseppunct}\relax
\EndOfBibitem
\bibitem[Hornak et~al.(2006)Hornak, Abel, Okur, Strockbine, Roitberg, and
  Simmerling]{hornak2006comparison}
Hornak,~V.; Abel,~R.; Okur,~A.; Strockbine,~B.; Roitberg,~A.; Simmerling,~C.
  \emph{Proteins: Struct., Funct., Bioinf.} \textbf{2006}, \emph{65},
  712--725\relax
\mciteBstWouldAddEndPuncttrue
\mciteSetBstMidEndSepPunct{\mcitedefaultmidpunct}
{\mcitedefaultendpunct}{\mcitedefaultseppunct}\relax
\EndOfBibitem
\bibitem[P{\'e}rez et~al.(2007)P{\'e}rez, March{\'a}n, Svozil, {\v{S}}poner,
  Cheatham~III, Laughton, and Orozco]{perez2007refinement}
P{\'e}rez,~A.; March{\'a}n,~I.; Svozil,~D.; {\v{S}}poner,~J.;
  Cheatham~III,~T.~E.; Laughton,~C.~A.; Orozco,~M. \emph{Biophys. J.}
  \textbf{2007}, \emph{92}, 3817--3829\relax
\mciteBstWouldAddEndPuncttrue
\mciteSetBstMidEndSepPunct{\mcitedefaultmidpunct}
{\mcitedefaultendpunct}{\mcitedefaultseppunct}\relax
\EndOfBibitem
\bibitem[Zgarbov{\'a} et~al.(2011)Zgarbov{\'a}, Otyepka, {\v{S}}poner,
  Ml{\'a}dek, Ban{\'a}{\v{s}}, Cheatham~III, and
  Jurecka]{zgarbova2011refinement}
Zgarbov{\'a},~M.; Otyepka,~M.; {\v{S}}poner,~J.; Ml{\'a}dek,~A.;
  Ban{\'a}{\v{s}},; Cheatham~III,~T.~E.; Jurecka,~P. \emph{J. Chem. Theory
  Comput.} \textbf{2011}, \emph{7}, 2886--2902\relax
\mciteBstWouldAddEndPuncttrue
\mciteSetBstMidEndSepPunct{\mcitedefaultmidpunct}
{\mcitedefaultendpunct}{\mcitedefaultseppunct}\relax
\EndOfBibitem
\bibitem[Bussi et~al.(2007)Bussi, Donadio, and Parrinello]{bussi2007canonical}
Bussi,~G.; Donadio,~D.; Parrinello,~M. \emph{J. Chem. Phys.} \textbf{2007},
  \emph{126}, 014101\relax
\mciteBstWouldAddEndPuncttrue
\mciteSetBstMidEndSepPunct{\mcitedefaultmidpunct}
{\mcitedefaultendpunct}{\mcitedefaultseppunct}\relax
\EndOfBibitem
\bibitem[Parrinello and Rahman(1981)Parrinello, and
  Rahman]{parrinello1981polymorphic}
Parrinello,~M.; Rahman,~A. \emph{J. Appl. Phys.} \textbf{1981}, \emph{52},
  7182--7190\relax
\mciteBstWouldAddEndPuncttrue
\mciteSetBstMidEndSepPunct{\mcitedefaultmidpunct}
{\mcitedefaultendpunct}{\mcitedefaultseppunct}\relax
\EndOfBibitem
\bibitem[Jorgensen et~al.(1983)Jorgensen, Chandrasekhar, Madura, Impey, and
  Klein]{jorgensen1983comparison}
Jorgensen,~W.~L.; Chandrasekhar,~J.; Madura,~J.~D.; Impey,~R.~W.; Klein,~M.~L.
  \emph{J. Chem. Phys.} \textbf{1983}, \emph{79}, 926--935\relax
\mciteBstWouldAddEndPuncttrue
\mciteSetBstMidEndSepPunct{\mcitedefaultmidpunct}
{\mcitedefaultendpunct}{\mcitedefaultseppunct}\relax
\EndOfBibitem
\bibitem[No{\'e} and Fischer(2008)No{\'e}, and Fischer]{noe2008transition}
No{\'e},~F.; Fischer,~S. \emph{Curr. Opin. Struct. Biol.} \textbf{2008},
  \emph{18}, 154--162\relax
\mciteBstWouldAddEndPuncttrue
\mciteSetBstMidEndSepPunct{\mcitedefaultmidpunct}
{\mcitedefaultendpunct}{\mcitedefaultseppunct}\relax
\EndOfBibitem
\bibitem[Pande et~al.(2010)Pande, Beauchamp, and Bowman]{pande2010everything}
Pande,~V.~S.; Beauchamp,~K.; Bowman,~G.~R. \emph{Methods} \textbf{2010},
  \emph{52}, 99--105\relax
\mciteBstWouldAddEndPuncttrue
\mciteSetBstMidEndSepPunct{\mcitedefaultmidpunct}
{\mcitedefaultendpunct}{\mcitedefaultseppunct}\relax
\EndOfBibitem
\bibitem[Chodera and No{\'e}(2014)Chodera, and No{\'e}]{chodera2014markov}
Chodera,~J.~D.; No{\'e},~F. \emph{Curr. Opin. Struct. Biol.} \textbf{2014},
  \emph{25}, 135--144\relax
\mciteBstWouldAddEndPuncttrue
\mciteSetBstMidEndSepPunct{\mcitedefaultmidpunct}
{\mcitedefaultendpunct}{\mcitedefaultseppunct}\relax
\EndOfBibitem
\bibitem[Prinz et~al.(2011)Prinz, Wu, Sarich, Keller, Senne, Held, Chodera,
  Sch{\"u}tte, and No{\'e}]{prinz2011markov}
Prinz,~J.-H.; Wu,~H.; Sarich,~M.; Keller,~B.; Senne,~M.; Held,~M.;
  Chodera,~J.~D.; Sch{\"u}tte,~C.; No{\'e},~F. \emph{J. Chem. Phys.}
  \textbf{2011}, \emph{134}, 174105\relax
\mciteBstWouldAddEndPuncttrue
\mciteSetBstMidEndSepPunct{\mcitedefaultmidpunct}
{\mcitedefaultendpunct}{\mcitedefaultseppunct}\relax
\EndOfBibitem
\bibitem[Bowman et~al.(2013)Bowman, Pande, and No{\'e}]{bowman2013introduction}
Bowman,~G.~R.; Pande,~V.~S.; No{\'e},~F. \emph{An introduction to markov state
  models and their application to long timescale molecular simulation};
  Springer Science \& Business Media: Heidelberg, Germany, 2013; Vol. 797\relax
\mciteBstWouldAddEndPuncttrue
\mciteSetBstMidEndSepPunct{\mcitedefaultmidpunct}
{\mcitedefaultendpunct}{\mcitedefaultseppunct}\relax
\EndOfBibitem
\bibitem[Bottaro et~al.(2014)Bottaro, Di~Palma, and Bussi]{bottaro2014role}
Bottaro,~S.; Di~Palma,~F.; Bussi,~G. \emph{Nucleic Acids Res.} \textbf{2014},
  \emph{42}, 13306--13314\relax
\mciteBstWouldAddEndPuncttrue
\mciteSetBstMidEndSepPunct{\mcitedefaultmidpunct}
{\mcitedefaultendpunct}{\mcitedefaultseppunct}\relax
\EndOfBibitem
\bibitem[Molgedey and Schuster(1994)Molgedey, and
  Schuster]{molgedey1994separation}
Molgedey,~L.; Schuster,~H.~G. \emph{Phys. Rev. Lett.} \textbf{1994}, \emph{72},
  3634\relax
\mciteBstWouldAddEndPuncttrue
\mciteSetBstMidEndSepPunct{\mcitedefaultmidpunct}
{\mcitedefaultendpunct}{\mcitedefaultseppunct}\relax
\EndOfBibitem
\bibitem[P{\'e}rez-Hern{\'a}ndez et~al.(2013)P{\'e}rez-Hern{\'a}ndez, Paul,
  Giorgino, De~Fabritiis, and No{\'e}]{perez2013identification}
P{\'e}rez-Hern{\'a}ndez,~G.; Paul,~F.; Giorgino,~T.; De~Fabritiis,~G.;
  No{\'e},~F. \emph{J. Chem. Phys.} \textbf{2013}, \emph{139}, 015102\relax
\mciteBstWouldAddEndPuncttrue
\mciteSetBstMidEndSepPunct{\mcitedefaultmidpunct}
{\mcitedefaultendpunct}{\mcitedefaultseppunct}\relax
\EndOfBibitem
\bibitem[Schwantes and Pande(2013)Schwantes, and
  Pande]{schwantes2013improvements}
Schwantes,~C.~R.; Pande,~V.~S. \emph{J. Chem. Theory Comput.} \textbf{2013},
  \emph{9}, 2000--2009\relax
\mciteBstWouldAddEndPuncttrue
\mciteSetBstMidEndSepPunct{\mcitedefaultmidpunct}
{\mcitedefaultendpunct}{\mcitedefaultseppunct}\relax
\EndOfBibitem
\bibitem[No{\'e} and Clementi(2015)No{\'e}, and Clementi]{noe2015kinetic}
No{\'e},~F.; Clementi,~C. \emph{J. Chem. Theory Comput.} \textbf{2015},
  \emph{11}, 5002--5011\relax
\mciteBstWouldAddEndPuncttrue
\mciteSetBstMidEndSepPunct{\mcitedefaultmidpunct}
{\mcitedefaultendpunct}{\mcitedefaultseppunct}\relax
\EndOfBibitem
\bibitem[MacQueen(1967)]{macqueen1967some}
MacQueen,~J. Some methods for classification and analysis of multivariate
  observations. 1967\relax
\mciteBstWouldAddEndPuncttrue
\mciteSetBstMidEndSepPunct{\mcitedefaultmidpunct}
{\mcitedefaultendpunct}{\mcitedefaultseppunct}\relax
\EndOfBibitem
\bibitem[Trendelkamp-Schroer et~al.(2015)Trendelkamp-Schroer, Wu, Paul, and
  No{\'e}]{trendelkamp2015estimation}
Trendelkamp-Schroer,~B.; Wu,~H.; Paul,~F.; No{\'e},~F. \emph{J. Chem. Phys.}
  \textbf{2015}, \emph{143}, 174101\relax
\mciteBstWouldAddEndPuncttrue
\mciteSetBstMidEndSepPunct{\mcitedefaultmidpunct}
{\mcitedefaultendpunct}{\mcitedefaultseppunct}\relax
\EndOfBibitem
\bibitem[Scherer et~al.(2015)Scherer, Trendelkamp-Schroer, Paul,
  Perez-Hernandez, Hoffmann, Plattner, Wehmeyer, Prinz, and
  No{\'e}]{scherer2015pyemma}
Scherer,~M.~K.; Trendelkamp-Schroer,~B.; Paul,~F.; Perez-Hernandez,~G.;
  Hoffmann,~M.; Plattner,~N.; Wehmeyer,~C.; Prinz,~J.-H.; No{\'e},~F. \emph{J.
  Chem. Theory Comput.} \textbf{2015}, \emph{11}, 5525--5542\relax
\mciteBstWouldAddEndPuncttrue
\mciteSetBstMidEndSepPunct{\mcitedefaultmidpunct}
{\mcitedefaultendpunct}{\mcitedefaultseppunct}\relax
\EndOfBibitem
\bibitem[Sch{\"o}lkopf et~al.(1997)Sch{\"o}lkopf, Smola, and
  M{\"u}ller]{scholkopf1997kernel}
Sch{\"o}lkopf,~B.; Smola,~A.; M{\"u}ller,~K.-R. In \emph{Proceedings of the
  Artificial Neural Networks --- ICANN'97: 7th International Conference,
  Lausanne, Switzerland, October 8--10, 1997}; Gerstner,~W., Germond,~A.,
  Hasler,~M., Nicoud,~J.-D., Eds.; Springer Berlin Heidelberg: Berlin,
  Heidelberg, 1997; pp 583--588\relax
\mciteBstWouldAddEndPuncttrue
\mciteSetBstMidEndSepPunct{\mcitedefaultmidpunct}
{\mcitedefaultendpunct}{\mcitedefaultseppunct}\relax
\EndOfBibitem
\bibitem[Cantor and Schimmel(1980)Cantor, and Schimmel]{cantor1980biophysical}
Cantor,~C.~R.; Schimmel,~P.~R. \emph{Biophysical Chemistry: Techniques for the
  Study of Biological Structure and Function. II}; WH Freeman: San Francisco,
  CA, USA, 1980; pp 402--403\relax
\mciteBstWouldAddEndPuncttrue
\mciteSetBstMidEndSepPunct{\mcitedefaultmidpunct}
{\mcitedefaultendpunct}{\mcitedefaultseppunct}\relax
\EndOfBibitem
\bibitem[No{\'e} et~al.(2013)No{\'e}, Wu, Prinz, and
  Plattner]{noe2013projected}
No{\'e},~F.; Wu,~H.; Prinz,~J.-H.; Plattner,~N. \emph{J. Chem. Phys.}
  \textbf{2013}, \emph{139}, 184114\relax
\mciteBstWouldAddEndPuncttrue
\mciteSetBstMidEndSepPunct{\mcitedefaultmidpunct}
{\mcitedefaultendpunct}{\mcitedefaultseppunct}\relax
\EndOfBibitem
\bibitem[No{\'e} et~al.(2011)No{\'e}, Doose, Daidone, L{\"o}llmann, Sauer,
  Chodera, and Smith]{noe2011dynamical}
No{\'e},~F.; Doose,~S.; Daidone,~I.; L{\"o}llmann,~M.; Sauer,~M.;
  Chodera,~J.~D.; Smith,~J.~C. \emph{Proc. Natl. Acad. Sci. U. S. A.}
  \textbf{2011}, \emph{108}, 4822--4827\relax
\mciteBstWouldAddEndPuncttrue
\mciteSetBstMidEndSepPunct{\mcitedefaultmidpunct}
{\mcitedefaultendpunct}{\mcitedefaultseppunct}\relax
\EndOfBibitem
\bibitem[Buchete and Hummer(2008)Buchete, and Hummer]{buchete2008coarse}
Buchete,~N.-V.; Hummer,~G. \emph{J. Phys. Chem. B} \textbf{2008}, \emph{112},
  6057--6069\relax
\mciteBstWouldAddEndPuncttrue
\mciteSetBstMidEndSepPunct{\mcitedefaultmidpunct}
{\mcitedefaultendpunct}{\mcitedefaultseppunct}\relax
\EndOfBibitem
\bibitem[D'Ascenzo et~al.(2016)D'Ascenzo, Leonarski, Vicens, and
  Auffinger]{dascenzo2016zdna}
D'Ascenzo,~L.; Leonarski,~F.; Vicens,~Q.; Auffinger,~P. \emph{Nucleic Acids
  Res.} \textbf{2016}, \emph{44}, 5944--5956\relax
\mciteBstWouldAddEndPuncttrue
\mciteSetBstMidEndSepPunct{\mcitedefaultmidpunct}
{\mcitedefaultendpunct}{\mcitedefaultseppunct}\relax
\EndOfBibitem
\bibitem[Bottaro et~al.(2016)Bottaro, Gil-Ley, and
  Bussi]{BottaroGil-LeyBussi2016}
Bottaro,~S.; Gil-Ley,~A.; Bussi,~G. \emph{Nucleic Acids Res.} \textbf{2016},
  \emph{44}, 5883--5891\relax
\mciteBstWouldAddEndPuncttrue
\mciteSetBstMidEndSepPunct{\mcitedefaultmidpunct}
{\mcitedefaultendpunct}{\mcitedefaultseppunct}\relax
\EndOfBibitem
\bibitem[Sokoloski et~al.(2011)Sokoloski, Godfrey, Dombrowski, and
  Bevilacqua]{sokoloski2011prevalence}
Sokoloski,~J.~E.; Godfrey,~S.~A.; Dombrowski,~S.~E.; Bevilacqua,~P.~C.
  \emph{RNA} \textbf{2011}, \emph{17}, 1775--1787\relax
\mciteBstWouldAddEndPuncttrue
\mciteSetBstMidEndSepPunct{\mcitedefaultmidpunct}
{\mcitedefaultendpunct}{\mcitedefaultseppunct}\relax
\EndOfBibitem
\bibitem[Zagrovic and Van~Gunsteren(2006)Zagrovic, and
  Van~Gunsteren]{zagrovic2006comparing}
Zagrovic,~B.; Van~Gunsteren,~W.~F. \emph{Proteins: Struct., Funct., Bioinf.}
  \textbf{2006}, \emph{63}, 210--218\relax
\mciteBstWouldAddEndPuncttrue
\mciteSetBstMidEndSepPunct{\mcitedefaultmidpunct}
{\mcitedefaultendpunct}{\mcitedefaultseppunct}\relax
\EndOfBibitem
\bibitem[Bottaro et~al.(2016)Bottaro, Ban{\'a}{\v{s}}, {\v{S}}poner, and
  Bussi]{bottaro2016free}
Bottaro,~S.; Ban{\'a}{\v{s}},~P.; {\v{S}}poner,~J.; Bussi,~G. \emph{J. Phys.
  Chem. Lett.} \textbf{2016}, \emph{7}, 4032--4038\relax
\mciteBstWouldAddEndPuncttrue
\mciteSetBstMidEndSepPunct{\mcitedefaultmidpunct}
{\mcitedefaultendpunct}{\mcitedefaultseppunct}\relax
\EndOfBibitem
\bibitem[Chen and Garc{\'\i}a(2013)Chen, and Garc{\'\i}a]{chen2013high}
Chen,~A.~A.; Garc{\'\i}a,~A.~E. \emph{Proc. Natl. Acad. Sci. U. S. A.}
  \textbf{2013}, \emph{110}, 16820--16825\relax
\mciteBstWouldAddEndPuncttrue
\mciteSetBstMidEndSepPunct{\mcitedefaultmidpunct}
{\mcitedefaultendpunct}{\mcitedefaultseppunct}\relax
\EndOfBibitem
\bibitem[Mills(1973)]{mills1973self}
Mills,~R. \emph{J. Phys. Chem.} \textbf{1973}, \emph{77}, 685--688\relax
\mciteBstWouldAddEndPuncttrue
\mciteSetBstMidEndSepPunct{\mcitedefaultmidpunct}
{\mcitedefaultendpunct}{\mcitedefaultseppunct}\relax
\EndOfBibitem
\bibitem[Mark and Nilsson(2001)Mark, and Nilsson]{mark2001structure}
Mark,~P.; Nilsson,~L. \emph{J. Phys. Chem. A} \textbf{2001}, \emph{105},
  9954--9960\relax
\mciteBstWouldAddEndPuncttrue
\mciteSetBstMidEndSepPunct{\mcitedefaultmidpunct}
{\mcitedefaultendpunct}{\mcitedefaultseppunct}\relax
\EndOfBibitem
\bibitem[Pinamonti et~al.(2015)Pinamonti, Bottaro, Micheletti, and
  Bussi]{pinamonti2015elastic}
Pinamonti,~G.; Bottaro,~S.; Micheletti,~C.; Bussi,~G. \emph{Nucleic Acids Res.}
  \textbf{2015}, \emph{43}, 7260--7269\relax
\mciteBstWouldAddEndPuncttrue
\mciteSetBstMidEndSepPunct{\mcitedefaultmidpunct}
{\mcitedefaultendpunct}{\mcitedefaultseppunct}\relax
\EndOfBibitem
\end{mcitethebibliography}
 \end{document}